\documentclass[fleqn,usenatbib]{mnras}

\usepackage{newtxtext,newtxmath}

\usepackage[T1]{fontenc}

\DeclareRobustCommand{\VAN}[3]{#2}
\let\VANthebibliography\thebibliography
\def\thebibliography{\DeclareRobustCommand{\VAN}[3]{##3}\VANthebibliography}

\usepackage{graphicx}	
\usepackage{amsmath}	
\usepackage{amssymb}	

\newcommand{\jj}{J0946+1006}	
\newcommand{\cgsflux}{erg\,s$^{-1}$\,cm$^{-2}$}
\newcommand{\z}{\phantom{0}}

\newcommand{\msun}{M$_\odot$}

\usepackage{color}

\newcommand{\rev}[1]{#1}

\defcitealias{2014MNRAS.443..969C}{CA14}
\newcommand{\ca}{\citetalias{2014MNRAS.443..969C}}

\title[A third source behind the Jackpot lens]{A Triple Rollover: A third multiply-imaged source at \boldmath{z}\,$\approx$\,6 behind the Jackpot gravitational lens}

\author[T.E. Collett and R.J. Smith]{
Thomas E. Collett$^{1}$ and  Russell J. Smith$^{2}$\thanks{E-mail:thomas.collett@port.ac.uk, russell.smith@durham.ac.uk. The authors contributed equally to this work. } 
\\
$^{1}$Institute of Cosmology and Gravitation, University of Portsmouth, Burnaby Rd, Portsmouth, PO1 3FX, United Kingdom\\
$^{2}$Centre for Extragalactic Astronomy, University of Durham, Durham DH1 3LE, United Kingdom\\
}

\date{Submitted to MNRAS, 2nd April 2020}

\pubyear{2020}

\begin{document}
\label{firstpage}
\pagerange{\pageref{firstpage}--\pageref{lastpage}}
\maketitle

\begin{abstract}
{\rev Using a five-hour adaptive-optics-assisted observation with MUSE,} 
we have identified a doubly-imaged Ly\,$\alpha$ source at redshift 5.975 behind the $z$\,=\,0.222 lens galaxy \jj\ (the `Jackpot'). 
The source separation implies an Einstein radius of $\sim$2.5\,arcsec.
Combined with the two previously-known Einstein rings in this lens (radii 1.4\,arcsec at $z$\,=\,0.609 and 2.1\,arcsec at $z$\,$\approx$\,2.4) 
this system is now a unique galaxy-scale triple-source-plane lens. 
We show that existing lensing models for \jj\ successfully map the two new observed images to a common point on the $z=5.975$ source plane.
The new source will provide further constraints on the mass distribution in the lens and in the two previously known sources. The third source also probes two new distance scaling factors which are sensitive to the cosmological parameters of the Universe. 
We show that detection of a new multiply imaged emission-line source is not unexpected in observations
of this depth; similar data for other known lenses should reveal a larger sample of multiple-image-plane systems for cosmography and other applications.
\end{abstract}




\section{Introduction}
\begin{figure*}
\vskip 0mm
\includegraphics[width=180mm]{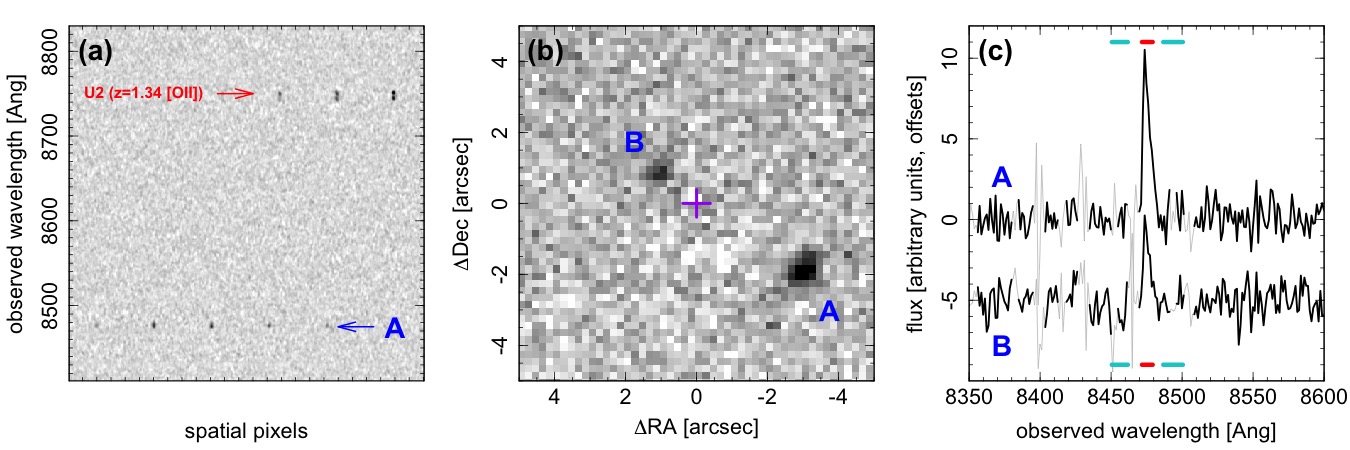} 
\includegraphics[width=180mm]{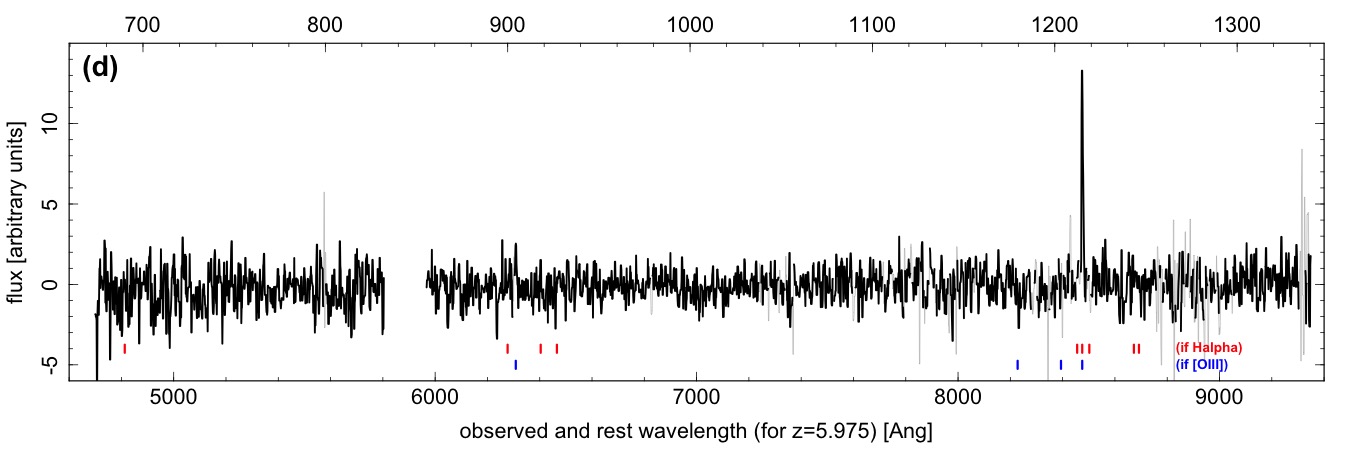} 
\vskip 0mm
\caption{Extracts from deep MUSE observations of \jj, highlighting the newly-discovered third lensed source. 
Panel (a) displays part of the unwrapped `inspection image' used to identify emission-line objects in the processed data cube. The object marked `A' is the feature of interest. (The [O\,{\sc ii}] emission at longer wavelengths is from a different, singly-imaged source projected 5.9\,arcsec from \jj, U2 in Table~\ref{tab:othersources}.)
Panel (b) shows a continuum-subtracted narrow-band image centred on the emission line at 8475\,\AA, showing the presence of an apparent counter-image, `B'.
Panel (c) presents the line profile for images A and B (extracted in apertures of diameter 0.8 and 0.5\,arcsec respectively), showing the asymmetric form characteristic of Ly\,$\alpha$, and demonstrating the close match in location and shape between A and B. Pixels affected by strong sky emission lines are marked in grey. The cyan and red bars indicate the wavelength ranges of the on- and off-band images used in panel (a). Panel (d) shows the full spectrum of image A. 
\rev{Red and blue tick marks show where other emission lines would be located if the bright line was [O\,{\sc iii}] 5007\AA\ or H\,$\alpha$, instead of Ly\,$\alpha$
}}
\label{fig:whammy}
\end{figure*}

The locations of images formed near to gravitational lenses depend on the mass distribution in the lens, the alignment between observer, lens and source and on angular-diameter distances between the observer, lens and source.
Galaxy-scale strong lensing thus finds application in a broad range of astrophysical enquiry, for example probing  the stellar initial mass function and dark matter (DM) halos in massive galaxies \citep[e.g.][]{2010ApJ...709.1195T,2010ApJ...721L.163A,2015MNRAS.449.3441S}, as well as measuring cosmological parameters \citep[e.g.][]{2012AJ....143..120O,2017MNRAS.465.4914B} and testing general relativity \citep{2018Sci...360.1342C}.

While hundreds of galaxy-scale lenses have now been discovered, there are very few known cases in which two sources at different redshifts
are multiply-imaged by the same foreground lens galaxy.  
Such systems are especially valuable for cosmological applications, 
since they allow the effects of the lensing potential and the distance ratios to be disentangled\footnote{Multiple source planes are ubiquitous in massive cluster-scale lenses, and attempts have been made to  exploit them for cosmological purposes \citep[e.g.][]{2010Sci...329..924J}, but the mass distribution in clusters is much more complex and irregular, which hinders this approach.}.
\cite{2012MNRAS.424.2864C} showed that competitive determinations of the dark energy equation-of-state parameter $w$ can be obtained with accurate measurements for just a handful of suitably-configured double-source-plane lens (DSPL) systems.
Moreover, DSPLs can provide constraints on the mass profile over a wider range of radii than usually spanned in single-image-plane systems, helping to break degeneracies between the stellar and DM mass components, as well as the cosmology.

The best-studied DSPL to date is the `Jackpot' lens \jj, which
was discovered serendipitously by \cite{2008ApJ...677.1046G}
as part of the SLACS (Sloan Lens ACS) survey \citep{2006ApJ...638..703B}.
The primary lens is a massive elliptical galaxy at $z_{\rm l}$\,=\,0.222, with velocity dispersion 284$\pm$24\,km\,s$^{-1}$ and effective radius 2.0\,arcsec (7.3\,kpc).
The SLACS spectroscopic lens search detected emission lines from a background source (s1) at $z_{\rm s1}$\,=\,0.609. {\it Hubble Space Telescope} (HST) observations showed that this 
source forms bright arcs with an Einstein radius 
1.4\,arcsec. 
The HST imaging also revealed a second system of arcs from a fainter source (s2), at a 
{\rev radius}
of   
2.1\,arcsec. 
The symmetric configuration 
shows that both of these sources are closely aligned with the mass centre.
The larger ring radius of s2 implies that it is more distant than s1, but only a photometric redshift estimate has yet been obtained, with 
$z_{\rm s2}$\,=\,2.41$^{+0.04}_{-0.21}$ \citep{2012ApJ...752..163S}.

The redshift configuration of \jj, with fairly small $z_{s1}$
compared to  $z_{l}$, but $z_{s2}$ substantially larger than both, is especially
favourable for deriving cosmological constraints \citep[see figure 4 of][]{2012MNRAS.424.2864C}.
By modelling the system with pixelised reconstructions of both source-planes,
\citet[][hereafter \ca]{2014MNRAS.443..969C}
derived $w$\,=\,--1.17$\pm$0.20 from this system alone, when
combined with a prior from cosmic microwave background measurements. 
Meanwhile \cite{2012ApJ...752..163S} have shown the utility of the double source plane for probing the mass profile of \jj, separating the stellar and DM components, with the inclusion of stellar kinematic data. Although a few further DSPLs have been
discovered since the Jackpot \citep{2009A&A...501..475T,2016ApJ...826L..19T,2019A&A...631A..40S}, none of these has the optimal combination of redshifts found in \jj.

In this paper, we use long-exposure integral field unit (IFU) observations to identify a third lensed source, s3, at a redshift of $z_{\rm s3}$\,=\,5.975 behind \jj, and discuss some of the applications of this system, which is now a unique triple-source-plane lens.
Section~\ref{sec:data} describes the discovery and observed characteristics of the new doubly-imaged background source. 
In Section~\ref{sec:modelling}, we develop a formalism to parameterise triple-source plane lenses, and demonstrate that the model of \ca\ successfully maps the two new images to a common origin in the distant source plane.
In Section~\ref{sec:outlook}, by considering the known population of emission-line source found in deep MUSE surveys, we show that the discovery of an additional lensed source in our data is not unexpected. 
Finally, Section~\ref{sec:concs} previews the further exploitation of \jj\ as a triple-source-plane lens, and highlights the possibility of `converting' other known 
lenses to DSPLs with future deep IFU observations.


\section{A third source plane in \jj}\label{sec:data}

We obtained deep IFU observations of \jj\ with MUSE  \citep[the Multi-Unit Spectroscopic Explorer;][]{2010SPIE.7735E..08B} on the ESO Very Large Telescope (Programme ID 0102.A-0950, PI: Collett), in 
February 2019. 
{\rev The observations were acquired in the wide-field mode, using the ground-layer adaptive optics system for improved image quality.}
We executed 7 observing blocks, each consisting of three 895\,sec exposures separated by small spatial dithers; 90-degree rotations were applied between the blocks.

We make use of the combined data cube generated through the standard observatory reduction pipeline and retrieved from the science archive. The total exposure time combined into the final data product is 5.2\,hours. The point-spread function has 0.5\,arcsec full-width at half maximum, estimated from a bright star in the field, at a wavelength of 8500\,\AA.
We processed the reduced data cube using methods developed from previous blind lens-search programmes \citep*{2015MNRAS.449.3441S,2020arXiv200207191C}. Briefly this scheme involves subtracting an elliptically-symmetric model for the spectrum of the primary lens, and then filtering out the residual low-order continuum light, and normalising to equalise the noise across the cube. Finally we apply a cleaning process based on principal components analysis, that helps to suppress the remaining sky-subtraction residuals. The resulting data cube is then unwrapped to a two-dimensional `inspection image' which can be visually examined to identify emission lines against a clean flat background. 

A small part of the unwrapped inspection image for 
\jj\ is shown in Figure~\ref{fig:whammy}a. 
Among the features identified by eye in this frame is an emission line at 8475\,\AA, marked `A' in the figure.
The spatial structure of this source can be seen in Figure~\ref{fig:whammy}b, which shows a net narrow-band image extracted from the original data cube, computed over a 5\,\AA\ interval centred at 8475\,\AA. 
(The off-band image was constructed from two 12.5\,\AA-wide intervals bracketing the line.)
The emission from image `A' is located 3.56\,arcsec from the centre of \jj\ and appears to be slightly extended along the tangential direction.
A counter-image, `B', is also visible, nearly diametrically opposite, at 1.35\,arcsec radius.
Figure~\ref{fig:whammy}c compares spectra extracted from these two regions, confirming a close match in both wavelength and line shape, with both having the characteristic asymmetric profile of Ly\,$\alpha$. Note that this can not be mistaken for the [O\,{\sc ii}] doublet, which would be clearly resolved (as seen in the other source visible in Figure~\ref{fig:whammy}a). The full spectrum of image A is shown in Figure~\ref{fig:whammy}d, confirming that no other lines are visible in the MUSE wavelength range. 
{ In particular we can rule out identifying the bright emission line as 
 [O\,{\sc iii}] 5007\,\AA\, given the absence of the corresponding 4959\,\AA\ line. Likewise, H\,$\alpha$ is excluded by absence of corresponding H\,$\beta$.}
We conclude that the emission is securely identified as a doubly-imaged $z$\,=\,5.975 Ly\,$\alpha$ source behind \jj, hereafter s3.


No continuum counterpart to s3 source is visible in HST imaging,
either in the SLACS F814W frame 
\citep[][2096\,sec exposure; Programme 10886; PI: Bolton]{2007ApJ...667..176G}
or in the F160W infrared data \citep[][2397\,sec; Programme 11202; PI: Koopmans]{2009ApJ...705.1099A}. The integrated line flux in image A is 7.8$\pm$0.5\,$\times$\,10$^{-18}$\,\cgsflux, which is fairly typical for $z$\,$\sim$\,6 Ly\,$\alpha$ emitters in deep MUSE observations 
\citep{2017A&A...608A...6D}. 
Such faint Ly\,$\alpha$ emitters frequently have no continuum counterpart even in HST observations much deeper 
than those available for \jj\ \citep{2015A&A...575A..75B}.

\begin{figure}
\vskip 0mm
\includegraphics[width=85mm]{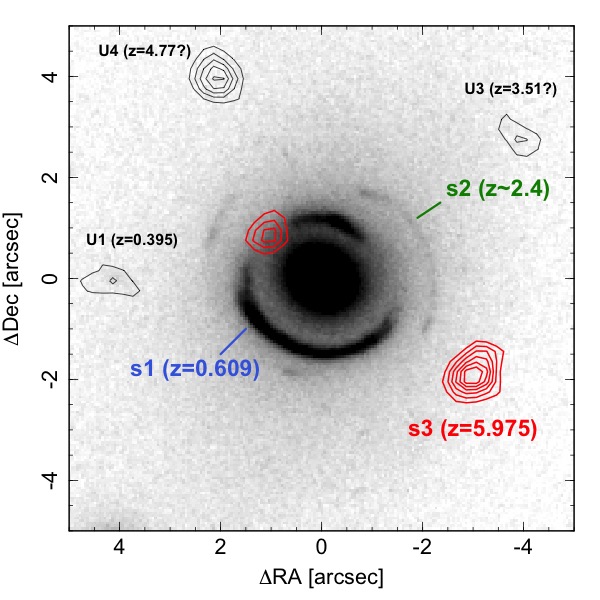} 
\vskip -2mm
\caption{Contours of the newly-discovered $z$\,=\,5.975 emission from MUSE (red), overlaid on the HST F814W image of \jj\ from SLACS. 
The thin black contours show three additional (singly-imaged)  emission-line sources also detected in the MUSE data.
}\label{fig:hstoverlay}
\end{figure}

Figure~\ref{fig:hstoverlay} shows the new MUSE source in the context of the previously-known lensing configuration from HST imaging. Note that with the mass model already secured by the s1 and s2 
arcs, the less perfect alignment of s3 is actually helpful, to probe the deflections 
at larger radius. Three additional, singly-imaged, emission-line sources at different redshifts are also seen in Figure~\ref{fig:hstoverlay}. The properties of all sources detected within 8\,arcsec are reported in Appendix~\ref{sec:sourcesearch}.

\begin{figure*}
\vskip 0mm
\includegraphics[width=\textwidth]{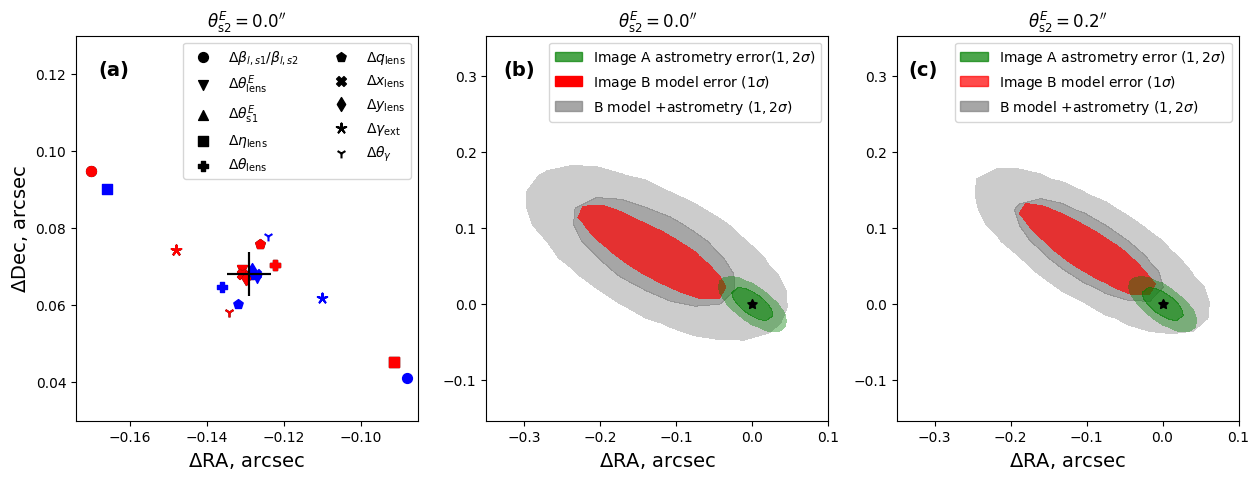} 
\vskip -1mm
\caption{
The delensed position of image B, relative to A, on the $z$\,=\,5.975 source plane. Panel (a) shows the offset for the model of \ca\ as published (large cross). The model, including deflections from the first source at $z$\,=\,0.609, maps  A and B to the same source-plane location within $\sim$0.15\,arcsec.  Red and blue points show the effect of $\pm$1$\sigma$ variations in individual parameters of the model; the leading contributions to the `model error' are the density profile slope of the primary lens, $\eta_{\rm lens}$, and the
cosmology-dependent factor, $\beta_{l,s1}/\beta_{l,s2}$ (this ratio is called $\beta$ in \ca).
Panel (b) shows that the positional offset is consistent with the combination of estimated astrometric errors and random errors on the \ca\ model fit.
Panel (c) shows the very minor effect of introducing mass on the second source plane at $z$\,=\,2.4, using an isothermal sphere corresponding to $\sigma$\,=\,150\,km\,s$^{-1}$, which far exceeds the likely mass of s2.
}
\label{fig:delens}
\end{figure*}

\section{Modelling the new source plane}\label{sec:modelling}

To understand fully the implications of the new $z$\,=\,5.975 source plane demands a comprehensive multiple-plane modelling analysis, sampling simultaneously over the mass distribution of the primary lens and s1 and s2 planes and the light distribution in all three sources, as well as the cosmological parameters (Collett et al., in preparation).
Here, we restrict our analysis to the simpler goal of testing whether the previously-published model of \ca\ can accommodate the new observation. 
The only additional element we introduce into the model is to test for the effect of mass in the s2 plane at $z$\,=\,2.4. 

\subsection{Formalism for triple source planes} 

The theory of multiplane lensing has been developed and applied to the double source plane case in previous work. Here we need to generalise this treatment to include the additional plane, which necessarily leads to some more complex notation.

For a single-source-plane lens, the lens equation can be written as
\begin{equation}
\label{eq:SSPL}
\mathbf{x}_s = \mathbf{x} - \boldsymbol{\alpha}(\mathbf{x}),
\end{equation}
where $\mathbf{x}_s$ is the angular position of the source on the source plane and  $\mathbf{x}$ is its position in the image plane. $\boldsymbol{\alpha}(\mathbf{x})$ is the deflection caused by the lens, scaled to angular coordinates where it is the difference between observed image position and the unlensed position on the source plane. $\boldsymbol{\alpha}$ is related to the physical deflection angle, ${\boldsymbol{\hat{\alpha}}}$, by
\begin{equation}
\boldsymbol{\alpha} = \frac{D_{l,s}}{D_{s}}\hat{\boldsymbol{\alpha}},
\end{equation}
where $D_{\mathrm{i,j}}$ are angular diameter distances between observer, lens and source. 

For a single lens plane with multiple source planes, the physical deflection angles, ${\boldsymbol{\hat{\alpha}}}$, depend only on where rays pass through the lens plane and not on which source plane they originate from. However, working with physical deflection angles requires angular diameter distances to be carried around with the equations. It is more convenient to define the reduced deflections for one source plane---conventionally the highest redshift source plane---and a set of scaling factors to convert the reduced deflections for the other source planes. The cosmological and redshift sensitivity is entirely encoded in these scaling factors enabling the lens modelling and cosmological parameter estimation to be decoupled into separate inference steps.

Equation \ref{eq:SSPL} can then be rescaled to yield 
\begin{equation}
\label{eq:SLPL}
\mathbf{x}_j = \mathbf{x} - \beta_{l,j} \boldsymbol{\alpha}_l(\mathbf{x}), 
\end{equation}
where $\mathbf{x}_j$ is the position on the $j$th source plane, $\boldsymbol{\alpha}_l$ 
is the angular deflection of the lens acting on rays from the highest-redshift source plane, and $\beta_{l,j}$ is the cosmological scaling factor:
\begin{equation}
\label{eq:beta}
\beta_{l,j} \equiv \frac{D_{l,j} D_{s}}{D_{j} D_{l,s}}, 
\end{equation}
with the $s$ referring to the highest redshift source, so that $\beta_{l,s}$ is 1.

In practice, there is also mass on the source planes. The equations for progressively more distant source planes are reached iteratively, as follows. Working outwards from the observer, rays reach the first source plane deflected by the primary lens:
\begin{equation}
\mathbf{x}_{s1}=\mathbf{x}-\beta_{l,s1}\boldsymbol{\alpha}_1(\mathbf{x}).
\end{equation}
On this plane they are deflected again by the first source. This deflection is a function of where the rays pass though the plane of s1:
\begin{equation}
\begin{aligned}
\mathbf{x}_{s2}&=\mathbf{x}-\beta_{l,s2}\boldsymbol{\alpha}_l(\mathbf{x})-\beta_{s1,s2}\boldsymbol{\alpha}_{s1}(\mathbf{x}_{s1})\\
&=\mathbf{x}-\beta_{l,s2}\boldsymbol{\alpha}_1(\mathbf{x})-\beta_{s1,s2}\boldsymbol{\alpha}_{s1}(\mathbf{x}-\beta_{l,s1}\boldsymbol{\alpha}_l(\mathbf{x}))
\end{aligned}
\end{equation}
The rays are there deflected once more before reaching the third and (currently for \jj) final source:
\begin{equation}\label{eqn:xs3}
\begin{aligned}
\mathbf{x}_{s3}
&=
\mathbf{x}-\boldsymbol{\alpha}_l(\mathbf{x})-\boldsymbol{\alpha}_{s1}(\mathbf{x}_2)-\boldsymbol{\alpha}_{s2}(\mathbf{x}_3)\\
&=
\mathbf{x}-\boldsymbol{\alpha}_l(\mathbf{x})-\boldsymbol{\alpha}_{s1}(\mathbf{x}-\beta_{l,s1}\boldsymbol{\alpha}_l(\mathbf{x})) \\
& \mathrel{\phantom{= x}} - \boldsymbol{\alpha}_{s2}(\mathbf{x}-\beta_{l,s2}\boldsymbol{\alpha}_1(\mathbf{x})-\beta_{s1,s2}\boldsymbol{\alpha}_{s1}(\mathbf{x}-\beta_{l,s1}\boldsymbol{\alpha}_l(\mathbf{x})))
\end{aligned}
\end{equation}
There are no cosmological scaling factors in the top version of this equation because of how the $\boldsymbol{\alpha}_{i}$ are defined: $\beta_{i,s3}$ is always 1.

\subsection{Testing the CA14 model} 
Given the formalism laid out in the previous subsection, we now investigate whether the model of \ca\ can reproduce the observed image locations of the new source.

Of the quantities entering into in Equation~\ref{eqn:xs3}, $\mathbf{x}$ is an observable and
$\beta_{l,s2}\boldsymbol{\alpha}_l(\mathbf{x})$, $\beta_{s1,s2}\boldsymbol{\alpha}_{s1}(\mathbf{x}_2)$ and $\beta_{l,s2}/\beta_{l,s1}$ are already constrained by the double source plane modelling in \ca. Respectively, these 
are called $\boldsymbol{\alpha}_l$, $\boldsymbol{\alpha}_{s1}$ and $\beta$ in \ca, but our discovery of a further source compels us to adopt a more general notation in this paper.

To project the model of \ca\ onto s3, we still need $\beta_{l,s1}$, $\beta_{s1,s2}$ and $\boldsymbol{\alpha}_{s2}(\mathbf{x}_s2)$. Since we are not attempting to constrain cosmology in this paper we assume $\beta_{l,s1}^{-1} = 1.49$ and $\beta_{s1,s2}^{-1} = 1.22$. These are derived assuming $z_{s2}=2.4$ and fixing a flat $\Lambda$CDM cosmology with $\Omega_\mathrm{M}=0.3$.

The \ca\ lensing model includes mass on the s1  
source plane, treated as a $\sigma$\,$\approx$\,100\,km\,s$^{-1}$ singular isothermal sphere (SIS), which contributes significantly to the deflections for s2, and also now for s3.  Equivalently, the model of the system should now include deflections from the s2 plane, acting on s3. In practice, however, the effect of s2 is expected to be quite modest.  \cite{2012ApJ...752..163S} quote an intrinsic (unlensed) magnitude of $H_{\rm AB}$\,=\,26.4, which at $z$\,$\approx$\,2.4 corresponds to a typical stellar mass 
of $\sim$10$^{8.5}$\,\msun, and an upper limit of 
$\sim$10$^{9.5}$\,\msun\ \citep[by comparison to][]{2015ApJ...801...97S}. Dynamical studies of low-mass galaxies at this redshift are limited, but suggest typical velocity scales equivalent to $\sigma$\,$\sim$\,80\,km\,s$^{-1}$ at $\sim$10$^{9.5}$\,\msun\ \citep{2019arXiv190209554P}, and extrapolating to lower mass suggests $\sigma$\,$\sim$\,60\,km\,s$^{-1}$ is a more plausible value for s3, and an Einstein radius of $\sim$0.1\,arcsec. 
With the inclusion of only a modest mass on the second source plane, the model of \ca\ 
should therefore be able to delens image A and B to a single self consistent location on the third source plane. For the mass of s2 we follow the method that \ca\ applied to s1: we place a SIS lens centred on the mean location of the brightest points in the four images of s2 delensed onto the second source plane. The only new free parameter in this model is then the Einstein radius of s2.



Given the above assumptions there are four remaining source of uncertainty in tracing images A and B back on to the source plane: the statistical uncertainties in the lens model of \ca; the astrometric uncertainty on the positions of images A and B; the Einstein radius of s2; and any systematic uncertainties in the assumed lens model. We illustrate in Figure~\ref{fig:delens} that the first three are sufficient to bring A and B to a single location in the third source plane. Neglecting astrometric uncertainties, the best fit model of \ca\ brings A and B to within 0.15\,arcsec of each other at $z$\,=\,5.975. This is reduced to 0.09\,arcsec by increasing the $\beta$ in \ca\ by $1 \sigma$ or to 0.10\,arcsec by decreasing the inferred density-profile slope of the primary lens by $1 \sigma$ (Figure~\ref{fig:delens}a). After accounting for the astrometric uncertainty in the observed image positions of A and B, the images are consistent with no offset between the unlensed source positions (Figure~\ref{fig:delens}b). Adding mass to s2 does not significantly change this conclusion: even if the velocity dispersion of s2 is $\sigma$\,=\,150\,km\,s$^{-1}$, it makes little difference to the offset between the unlensed positions of A and B (Figure~\ref{fig:delens}c).

The fact that the model of \ca\  can bring images A and B to a single point means that there is no strong indication of systematics in the lens model or for deviations from a flat $\Lambda$CDM cosmology with $\Omega_\mathrm{M}=0.3$. 

{\rev{This consistency, for a source plane at $z$\,=\,5.975, provides support for 
a `gravimetric redshift' of $z_{s3}$\,$\approx$\,6, validating the identification of the emission line in A and B as Ly\,$\alpha$. If we instead allow $z_{s3}$ to be a free parameter, we find that the \ca\ model strongly prefers a high source redshift. We estimate the gravimetric redshift using an approximate Bayesian computation approach: for each of 1000 realisations of the \ca\ model (consistent with the \ca\ quoted errors), we forward model 1000 realisations of the positions of image A and B (consistent with the astrometric errors) onto each of 100 source planes between redshift 4 and 8. We find that none brings A and B closer than 0.2 arcseconds for source planes with $z_{s3}$\,$<$\,4.7, whereas there are models that bring A and B close together for redshifts above $4.7$. Given the bright emission line at 8475\,\AA, only Ly\,$\alpha$ at $z$\,=\,5.975 is a realistic solution.
 This approach of using a lens model to confirm an otherwise tentative source redshift was previously used in \cite{2013ApJ...762...32C}.
}}

\section{Likelihood of finding the new source}\label{sec:outlook}




 In this section, we assess the likelihood of having discovered another emission-line source behind \jj, using the  published empirical source counts from deep MUSE observations, together with a simplified treatment of the
 lensing properties of the system.
 
 The deepest published MUSE source catalogue is derived from the central 1.15\,arcmin$^2$ region of the Hubble Ultra Deep Field (HUDF), with a total exposure time of 31\,h  and FWHM $\sim$0.6\,arcsec in the red \citep{2017A&A...608A...1B}. From this region, 
 \cite{2017A&A...608A...2I} report $\sim$300 sources with fluxes down to 10$^{-18}$\,\cgsflux. Hence, to this intrinsic depth, the $\sim$10\,arcsec$^2$ multiply-imaged area of the source plane behind a massive elliptical galaxy should provide an average $\sim$1 detectable background galaxy. Although at 5.2\,h, our exposure time is shorter, this is more than compensated by the typical lensing amplification factor of $\sim$3 in the multiply-imaged regime, and aided by our slightly better image quality.
 
 To estimate more rigorously the expected yield of multiply-imaged sources as a function of the observed flux limit, we consider the contribution that each source from the HUDF catalogue would make to the population behind \jj. We first determine the Einstein radius for the given source redshift, assuming a SIS with $\sigma$\,=\,287\,km\,s$^{-1}$ at $z$\,=\,0.222 for the primary lens, augmented by a second SIS of $\sigma$\,=\,100\,km\,s$^{-1}$ at $z$\,=\,0.609, representing s1.
 We then use the catalogue source flux
 and the relationship between magnification and radius 
 for a SIS lens to determine the effective area inside which the source would exceed our flux limit, after lensing\footnote{Note however that the magnification properties of compound lenses can be significantly more complex than assumed here \citep{2016MNRAS.456.2210C}.}. The area is capped at the Einstein radius, since we wish to count only the multiply-imaged sources. 
 Comparing this area to the full area of the HUDF catalogue gives the contribution of this source to the yield, and we sum over all entries to determine the total expected number of lensed sources. 

\begin{figure}
\vskip 0mm
\includegraphics[width=85mm]{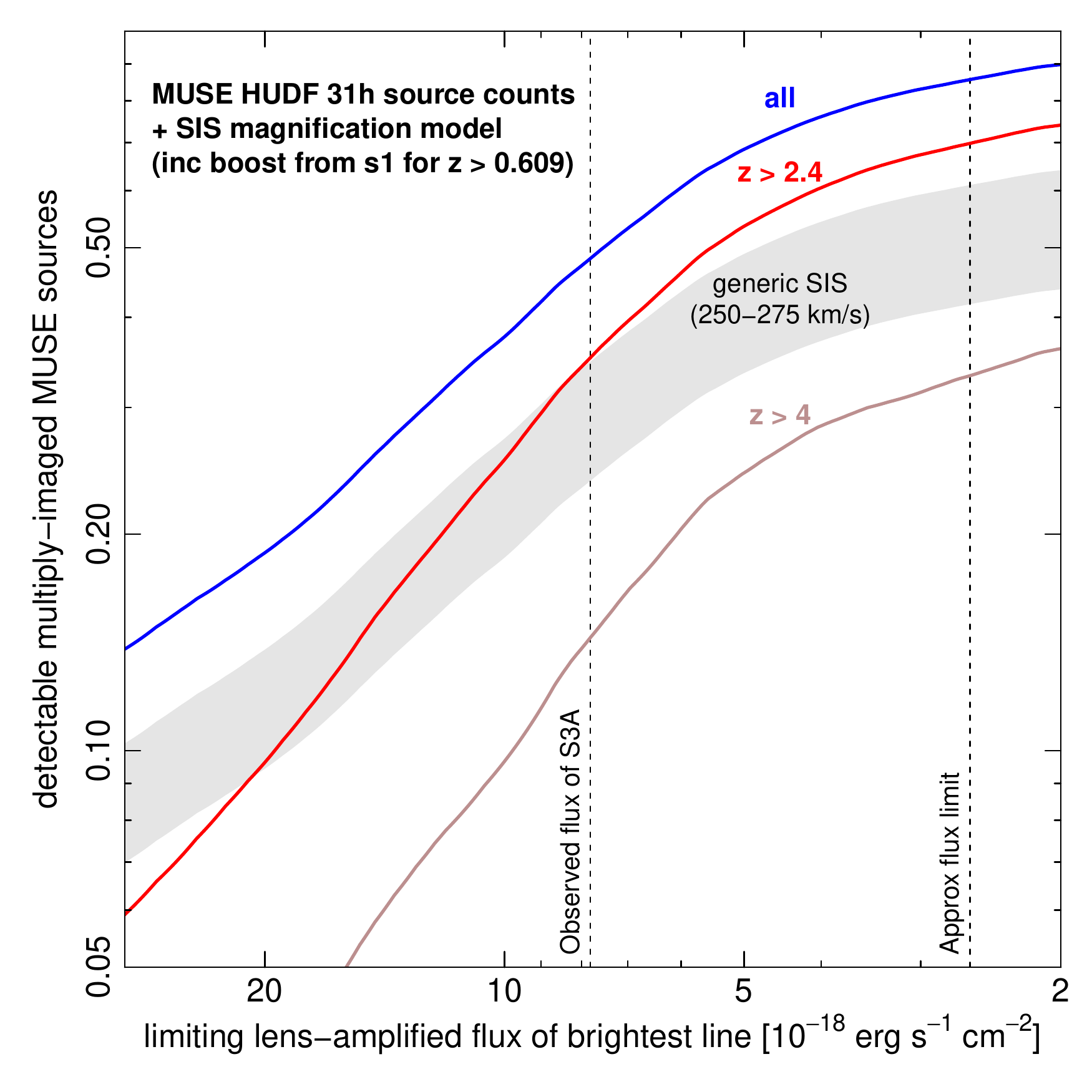} 
\vskip -1mm
\caption{Predicted number of multiply-imaged MUSE emission-line sources behind \jj\ as a function of limiting flux. The curves are calculated from empirical source counts from Inami et al. (2017).
The lensing cross-sections and amplifications are determined from a singular isothermal sphere model, including 
a contribution from the bright source s1. 
No correction has been made for increased noise due to the foreground continuum from the lens. 
The grey band shows the equivalent calculation for a typical single-plane lens at $z$\,$\approx$\,0.2 with $\sigma$ in the range 250--275\,km\,s$^{-1}$.
}\label{fig:boostedcounts}
\end{figure}

Figure~\ref{fig:boostedcounts} shows the results of these calculations as a function of flux limit, with three different redshift selections. We find that we should expect to detect a multiply-imaged source as bright as s3 in $\sim$50 per cent of realisations, although few of these should be at such high redshift.
To the approximate limit of our observations, which we estimate to be a factor of $\sim$3 fainter than s3 for reliable detection, the number of detectable lensed sources per realisation approaches unity. An important caveat to this is that identifying a lens requires detecting not only the most amplified source but also a secure counter-image, which might be substantially fainter. 
Nevertheless, we conclude that our discovery of a further source behind \jj\ should not be surprising, given the characteristics of our data.




\section{Summary and outlook}\label{sec:concs}

We have presented the discovery of a third multiply-imaged source behind the `Jackpot' lens galaxy \jj.
The new source plane at $z$\,=\,5.975 opens up sensitivity to further combinations of cosmological distances which enter into the lens equation, and hence leads to improved constraints on cosmological parameters. Moreover the new source probes new parts of the image plane, with image A almost doubling the radial distance from the centre of the primary lens compared to the images of s1 and s2. This extra information will help to fit more realistic 
density profiles distinguishing between stellar and dark matter components. In turn, this helps break degeneracies with the density profile, and hence further hone the cosmological sensitivity. The new source also enables constraints on the mass of s2 and on the mass profile of s1. Currently s1 has been assumed to be an SIS in the model of \ca\, however it is clear from the reconstruction that the light profile of s1 is elongated and clumpy. Both s1 and s2 are too low mass to be detectable strong lenses in their own right - analysing their perturbative effect on a compound lens is perhaps the only way to directly measure the relationship between the mass and light distributions of such small isolated galaxies.

\begin{figure}
\vskip 0mm
\includegraphics[width=85mm]{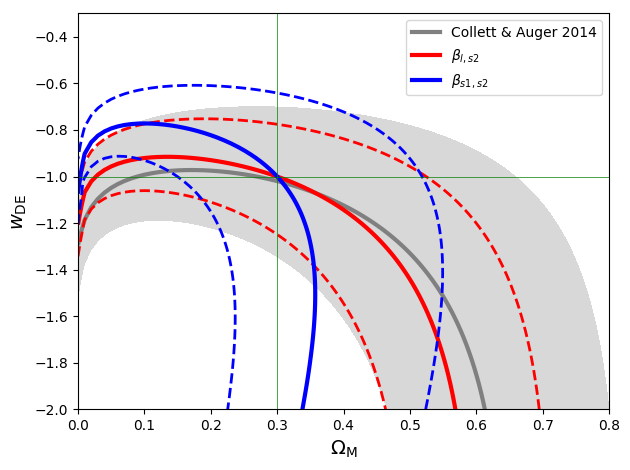} 
\vskip -1mm
\caption{Projected sensitivity of the cosmological scale factors to the cosmological parameters in a flat $w$CDM cosmology. The grey band shows the best fit and 68\% confidence region constraints from \ca. Red contours indicate the best fit and 68\% confidence region assuming a constraint on $\beta_{l,s2}$ can be derived with the same 1.1\% fractional precision as in \ca. The blue contours assume the same but for $\beta_{s1,s2}$. The green cross shows the input assumed true cosmology with $\Omega_{\mathrm{M}}=0.3$ and $w=-1$. }\label{fig:cornerplot}
\end{figure}

As shown in Section~\ref{sec:modelling}, the addition of a third source plane redshift opens up two new cosmological distance ratios:
\begin{equation}
\label{eq:newbeta}
\beta_{l,s2} \equiv \frac{D_{l,s2} D_{s3}}{D_{s2} D_{l,s3}} \text{\ , \ \ and \ \ } \beta_{s1,s2} \equiv \frac{D_{s1,s2} D_{s3}}{D_{s2} D_{s1,s3}} 
\end{equation}
Measuring a single $\beta$ provides a degenerate measurement of cosmological parameters for models beyond flat $\Lambda$CDM. \ca\ broke this degeneracy using Cosmic Microwave Background constraints, but combining multiple $\beta$ ratios will break this degeneracy without recourse to an external dataset. \cite{2012MNRAS.424.2864C} showed that the longer lever-arm to s3 gives greater sensitivity of $\beta_{l,s2}$ to $w$ compared to the $\beta$ measured in \ca\footnote{This distance ratio is also secured wholly with  spectroscopic redshifts, bypassing any concern over using the photometric estimate for s2, though the uncertainty is formally small.}.
Measuring $\beta_{s1,s2}$ adds a higher redshift lens plane which would be more constraining of evolving dark energy models \citep{2016PhRvD..94h3510L}.
In Figure \ref{fig:cornerplot} we show the projected constraints on $w$ and $\Omega_{\mathrm{M}}$ in a flat $w$CDM cosmology, assuming each $\beta$ can be measured to the same precision as in \ca. With current data $\beta_{s1,s2}$ is likely to be almost unconstrained since $\boldsymbol{\alpha}_{s1}$ is already small. However matching the precision of \ca\ for $\beta_{l,s2}$ is plausible with existing data. The cosmological inference in \ca\ should also be improved by modeling the third source since this extra source further constrains the other parameters of the lens model.

Completing the modelling for \jj\ will require a simultaneous reconstruction of all three sources (an approach pioneered by \citealt{2003ApJ...590..673W}) and of the spatially resolved kinematics of the primary lens and s1 (Collett et al in prep). Our MUSE data were originally obtained to provide this kinematic data: the discovery of a third source plane is a serendipitous---though not unforeseen---bonus. Simultaneously fitting the three sources and two sets of kinematics will enable an exploration of the full range of mass profiles that can reconstruct the observed data. We must also include perturbative line of sight effects \citep{2009A&A...499...31H,2013MNRAS.432..679C,2017JCAP...04..049B,2017ApJ...836..141M,2017MNRAS.467.4220R}, since ignoring them will introduce a systematic on the cosmological inference comparable to the expected statistical uncertainties. Accounting for all these complexities is far beyond the scope of the present paper.

Looking beyond applications of \jj\  
specifically, our discovery suggests that faint multiply imaged line emitters are common behind massive galaxies.
The calculations in Figure~\ref{fig:boostedcounts} are broadly applicable to any similarly massive galaxy, with the overall yield scaling with $\sigma^4$. (The system-specific boost to the counts from s1 is only $\sim$20 per cent.)  
Hence, a programme of deep IFU observations for more known lenses could reveal additional faint lensed sources in other systems, and hence establish a meaningful {\it sample} of double source-plane lenses for cosmology and other applications\footnote{\rev In MUSE data acquired for 
other purposes,  
we have already identified a quadruply-imaged second background source behind J1143--0144. The redshift configuration of this system 
($z_{\rm l}, z_{\rm s1}, z_{\rm s2}$)\,=\,(0.10, 0.40, 0.79) is much less sensitive to cosmology than that of \jj.}.
These arguments also apply to galaxies that are not known to be lenses at all. \citet{2015MNRAS.449.3441S} remarked that `strongly lensed background line emitters could be detected for any chosen massive elliptical', given IFU observations of sufficient depth. In this context, note that the sensitivity of our current MUSE data should be reached in less than 15\,minutes integration with the 39m E-ELT. With deep IFU observations from such telescopes, any massive galaxy can be converted into a strong lens, and any known lens can be converted into a multiple-source-plane system. 

\section*{Acknowledgements}
TEC was supported by a Royal Astronomical Society Fellowship and the University of Portsmouth Dennis Sciama Fellowship.
RJS was supported by the Science and Technology Facilities Council through the Durham Astronomy Consolidated Grant 2017--2020 (ST/P000541/1). 
This work is based on observations collected at the European Organisation for Astronomical Research in the Southern Hemisphere under ESO programme 102.A-0950.

\section*{Data Availability}
Data used in this paper are publicly available in the ESO (\url{archive.eso.org}) and HST (\url{hla.stsci.edu}) archives. 




\bibliographystyle{mnras}
\bibliography{triplewhammy} 

\appendix

\section{Source search}\label{sec:sourcesearch}

For completeness, we have searched the inspection image for additional emission-line sources projected close to \jj.  Six further sources  detected within 8\,arcsec are summarized in Table~\ref{tab:othersources}. 

Among these emitters, U4 is located closest to the multiply-imaged region at its corresponding source redshift. U4 has a single relatively bright emission line, which is likely to be Ly\,$\alpha$ at $z$\,=\,4.77. The model of \ca\ gives a less than 1 in 1000 chance that U4 is multiply imaged. Even if it is multiply imaged, any counter-image would be very faint and close to the lens centre, and would not be detectable in the present data. In principle, a secure upper limit on the counter-image flux can provide further (weak) constraints on viable lensing models, in a similar vein to \cite*{2018MNRAS.481.2115S}.

Note that none of the detected line emitters correspond to the s2 Einstein ring. We have also inspected a spectrum extracted from a mask centred on the s2 arcs, but neither any emission nor any absorption features could be identified. This is not unexpected given the photometric redshift of \cite{2012ApJ...752..163S} and gravimetric redshift of \ca, as the MUSE data would cover $\sim$1400--2800\,\AA\ in the rest frame, where only very weak features are present. 
To secure a spectroscopic redshift for this source will likely call for IFU observations in the near infrared (for the rest-frame-optical lines) or in the blue, targeting Ly\,$\alpha$ at $\sim$4100\,\AA. 

\begin{table}
\caption{All MUSE-detected emission-line sources within 8\,arcsec from \jj. The properties of the doubly-imaged source are given first, followed by six singly-imaged sources U1--U6. Only the first two of these have completely secure redshifts. Sources U3--U6 have only a single line detection, and we tabulate the corresponding redshift assuming it is Ly\,$\alpha$. However unlike in the images of s3 (A and B), the line cannot be confidently identified from its profile in these faint sources.
Positions are in arcsec relative to the lens galaxy centroid, and `rad' indicates radius in arcsec. Line fluxes are in 10$^{-18}$\,\cgsflux.
}\label{tab:othersources}
    \centering
    \begin{tabular}{lcccccc}
    \hline
    ID & $\Delta$RA & $\Delta$Dec & rad & flux & line & $z$ \\
    \hline
A & +3.00$\pm$0.02 & --1.91$\pm$0.02 & 3.56 & \z7.8$\pm$0.5 & Ly\,$\alpha$ & 5.975 \\
B & --1.03$\pm$0.05 & +0.87$\pm$0.04 & 1.35 & \z4.7$\pm$0.6 & Ly\,$\alpha$ & 5.975 \\
\hline
U1 & --4.05$\pm$0.06 & --0.00$\pm$0.06 & 4.05 & \z1.5$\pm$0.2 & [O\,{\sc iii}] & 0.395 \\
U2 & +5.71$\pm$0.02 & --1.24$\pm$0.02 & 5.85 & 13.3$\pm$0.7 & [O\,{\sc ii}] & 1.347 \\
U3 & +4.01$\pm$0.07 & +2.76$\pm$0.04 & 4.87 & \z2.1$\pm$0.3 &   Ly\,$\alpha$? & 3.508 \\
U4 & --2.09$\pm$0.03 & +3.96$\pm$0.03 & 4.48 & \z4.6$\pm$0.2 &   Ly\,$\alpha$? & 4.768 \\
U5 & +3.56$\pm$0.07 & +5.88$\pm$0.09 & 6.87 & \z2.3$\pm$0.3  &   Ly\,$\alpha$? & 5.046 \\
U6 & +3.24$\pm$0.05 & --5.22$\pm$0.04 & 6.15 & \z2.2$\pm$0.2 &   Ly\,$\alpha$? & 5.723 \\
\hline

\end{tabular}
\end{table}
\bsp	
\label{lastpage}
\end{document}